\documentclass[prl,twocolumn,superscriptaddress,showpacs]{revtex4}
\usepackage{amsmath,amssymb,graphicx,color,bm,epstopdf}

\begin{document}
\title{Nonlinear terahertz emission in semiconductor microcavities}

\author{I. G. Savenko}
 \affiliation{Science Institute, University of Iceland, Dunhagi-3, IS-107, Reykjavik, Iceland}
\affiliation{Academic University - Nanotechnology Research and
Education Centre, Khlopina 8/3, 195220, St.Petersburg, Russia}

\author{I. A. Shelykh}
 \affiliation{Science Institute, University of Iceland, Dunhagi-3, IS-107, Reykjavik, Iceland}
 \affiliation{International Institute of Physics, Av. Odilon Gomes de Lima, 1772, Capim Macio, 59078-400, Natal, Brazil}

\author{M. A. Kaliteevski}
\affiliation{Academic University - Nanotechnology Research and
Education Centre, Khlopina 8/3, 195220, St.Petersburg, Russia}
\affiliation{Ioffe Physical-Technical Institute, Polytekhnicheskaya
26, 194021, St.Petersburg, Russia}

\date{\today}

\begin{abstract}

We consider the nonlinear terahertz emission by the system of cavity
polaritons in the regime of polariton lasing. To account for the
quantum nature of terahertz- polariton coupling we use Lindblad
master equation approach and demonstrate that quantum microcavities
reveal rich variety of the nonlinear phenomena in terahertz range,
including bistability, short THz pulse generation and THz switching.
\end{abstract}


\maketitle


\emph{Introduction}. THz band remains the last region of
electromagnetic spectrum which does not have wide application in
modern technology due to lack of solid state source of THz
radiation which is compact, reliable and scalable [\onlinecite{Davies}].
Fundamental objection preventing realization of such source is small
rate of spontaneous emission of the THz photons. According to Fermi Golden rule this rate is about tens of inverse milliseconds,
while lifetime of the photoexcited carrier typically
lies in picosecond range due to the efficient interaction with
phonons \cite{Duc,Doan}. Spontaneous emission rate can be increased
by application of Purcell effect when emitter of THz is placed in
cavity for THz mode \cite{Todorov1,Chassagneux1}, but even in this
case cryogenic temperatures are required to provide quantum
efficiency of the order about one percent for typical quantum cascade
structure.

Recently it was proposed that the rate of spontaneous emission for THz
photons  can be additionally increased by bosonic stimulation if
radiative  transition occurs into a condensate state
\cite{Kavokin1}.  One example is a transition between upper and
lower polariton branches in semiconductor  microcavity in the regime
polariton lasing. Unfortunately, the radiative transition accompanied by emission of
THz photon between upper and lower polariton modes is forbidden,  since these states have the same parity.
Nevertheless, such transition becomes possible if upper polariton state is
mixed with exciton state of  different parity. Amplification of
spontaneous emission by Purcell effect together with bosonic
stimulation  increase the rate of spontaneous emission by several
orders of magnitude, making it comparable  with the rate of scattering
with acoustic phonon. Consequently, effective emission of THz radiation can occur
\cite{Kavokin1}.

It is well known that strong polariton-polariton interactions in
microcavities make it possible to observe pronounced nonlinear
effects for the intensities of the pump orders of
magnitude smaller than in other nonlinear optical  systems. Among
them are polariton superfluidity \cite{Amo}, bistability and
multistability \cite{Baas,Gippius}, soliton formation \cite{Egorov}
and others. One can expect that polariton-polariton interactions
will as well strongly affect the process of THz emission.
The quasiclassical approach based on Boltzmann equations,
used in Ref.\onlinecite{Kavokin1} cannot provide a correct account
of a coherent interaction of THz photons and polaritons, and can not be used for satisfactory description of nonlinearities in the considered system. The development of more exact quantum formalism is thus  needed. This paper is
aimed at building such a formalism, which accounts for the following
physical processes: coherent polariton-THz photon interaction,
polariton- polariton interaction leading to the blueshift of the
polariton modes and coupling of the polaritons with acoustic
phonons. The development of such description is timely in light of
intensive studies of ultrastrong light-matter coupling
[\onlinecite{Todorov2},\onlinecite{Gunter}], single cycle THz
generation \cite{Junginger,Todorov3}, intersubband cavity polariton physics
\cite{Todorov3,Ciuti} and control of the phase of THz radiation
 in both inorganic and organic structures \cite{Oustinov,Swoboda}.

\begin{figure}[!htb]
\includegraphics[width=1.0\linewidth]{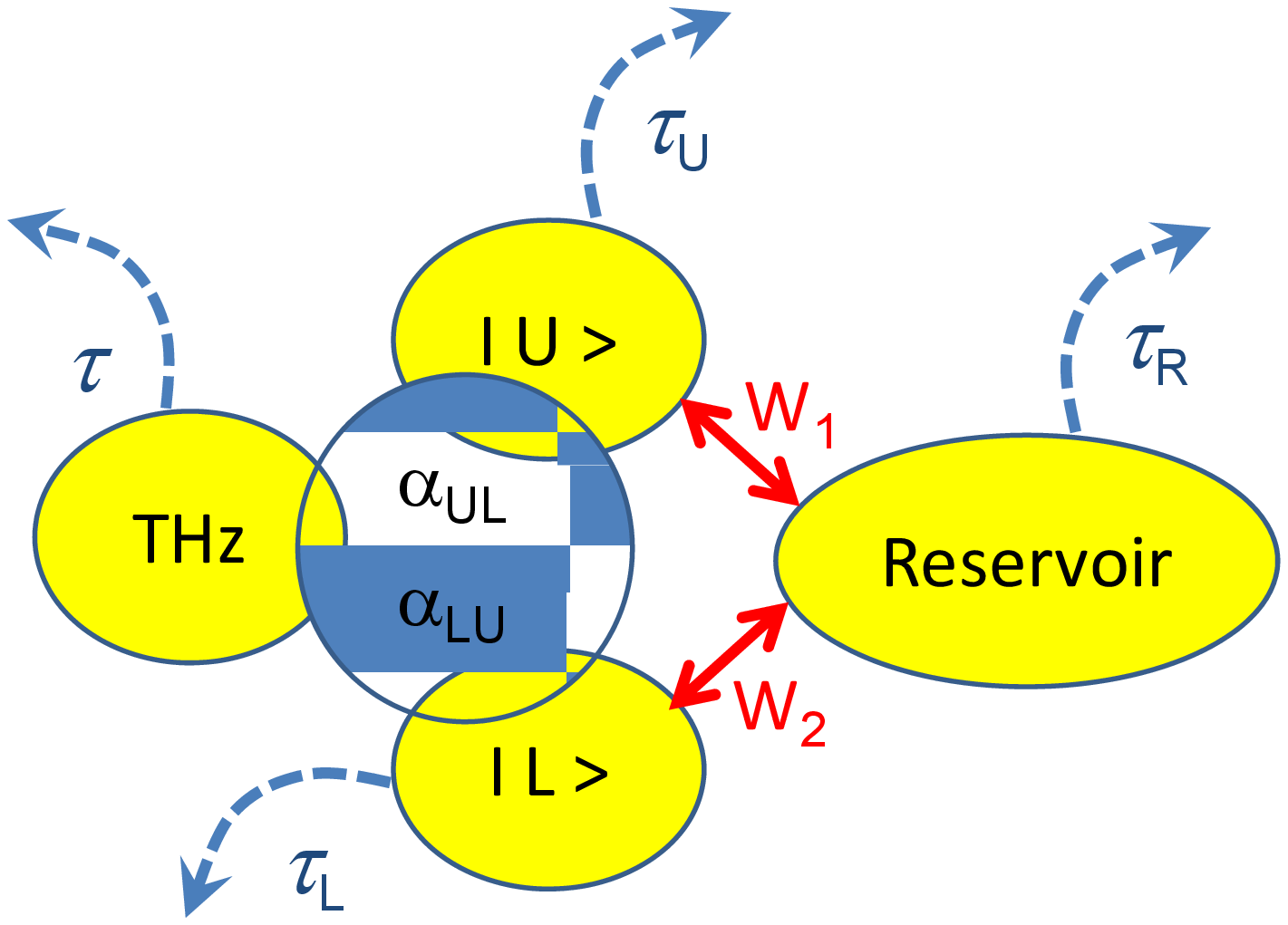}
\caption{The scheme of transitions in THz emitting cavity.
The upper polariton is mixed with dark exciton state due to the
application of the gate voltage $V_g$. The radiative transition
between the upper hybrid state $|U\rangle$ and lower polariton
state $|L\rangle$ thus becomes possible. Upper hybrid and lower
polariton states are also coupled with an incoherent reservoir of
the polaritons via phonon-assisted process.} \label{fig:1}
\end{figure}

\emph{Formalism}. We consider a model system consisting of a lower
polariton state with the energy $\epsilon_L$, upper hybrid state with the
energy $\epsilon_U$, THz cavity mode with the energy $\epsilon_T$ and
incoherent polariton reservoir coupled with upper and lower
polariton states via phonon-assisted process (see Fig.\ref{fig:1}).

The Hamiltonial of the system written in terms of the operators of
secondary quantization for upper polaritons ($a_U,a_U^+$), lower
polaritons ($a_L,a_L^+$), TH photons ($c,c^+$), reservoir states
($a_{R\textbf{k}},a^+_{R\textbf{k}}$) and acoustic phonons
$b_{\textbf{k}},b^+_{\textbf{k}}$ can be represented as a sum of
four terms:
\begin{eqnarray}
H=H_0+H_{pol-pol}+H_{T}+H_R
\end{eqnarray}

The first term
\begin{equation}
H_0=\epsilon_La^+_La_L+\epsilon_Ua^+_Ua_U+\epsilon_Tc^+c+\sum_k\epsilon_\textbf{k}a^+_{R\textbf{k}}a_{R\textbf{k}}
\end{equation}
corresponds to the energy of uncoupled upper and lower polaritonic
states, THz mode and polariton reservoir.

The second term
\begin{eqnarray}
H_{pol-pol}=U_{LL}a^+_La^+_La_La_L+U_{UU}a^+_Ua^+_Ua_Ua_U+\\
\nonumber+2U_{UL}a^+_Ua_Ua^+_La_L+\sum_k\left(U_{LR}a^+_La_L+U_{UR}a^+_Ua_U\right)a^+_{R\textbf{k}}a_{R\textbf{k}}
\end{eqnarray}
describes polariton- polariton interaction. The interaction
constants can be estimated as $U_{ij}=X_i^2X_j^2U$, where $X_i$ are
Hopfield coefficients giving the percentage of the exciton
fraction in the polariton states. $X_U$ and $X_L$ are determined by
cavity geometry, and we took $X_R=1$ supposing that the reservoir is
purely excitonic. The matrix element of the exciton- exciton
scattering can be estimated as $U\approx 6E_Ba_B^2/S$ with $E_B$ and
$a_B$ being the exciton binding energy and Bohr radius
respectively, and $S$ the area of the system
\cite{CiutiMatrElement}.

The third term
\begin{eqnarray}
H_T=V_T(a_U^+a_Lc+a_Ua_L^+c^+)
\end{eqnarray}
describes radiative THz transion between upper and lower polariton
states. The matrix element of the THz emission can be estimated
using a standard formula for the coupling constant of the dipole
transition with confined electromagnetic mode,
$V_T=\omega^2d\sqrt{\hbar n/2\pi^3\epsilon_0c^3}$, where $d$ is
matrix element of the radiative transition and n- refrective index
of the terahertz cavity (See e.g. Ref.\onlinecite{Scully}).

Interaction between upper and lower polariton states and incoherent
reservoir is described by the fourth term:
\begin{eqnarray}
H_R=H_R^++H_R^-=D_1\sum_{\textbf{k}}\left(a_Ua_{R\textbf{k}}^+b_{\textbf{k}}^++a_U^+a_{R\textbf{k}}b_{\textbf{k}}\right)+ \\
\nonumber+D_2\sum_{\textbf{k}}\left(a_La_{R\textbf{k}}^+b_{\textbf{k}}+a_L^+a_{R\textbf{k}}b_{\textbf{k}}^+\right)
\end{eqnarray}
where  $b_{\textbf{k}}^+$ and $b_{\textbf{k}}$ denote operators of
creation and annihilation of phonons with wavevector $\textbf{k}$,
$D_1$ and $D_2$ are the polariton-phonon interaction constants.

Keeping in mind that interactions described by $H_0,H_{pol-pol},H_T$
are of coherent nature, while phonon assisted interactions ($H_R$)
with reservoir destroy coherences the dynamics of the density matrix
of system $\rho$ is described by Lindblad master equation,
analogical to those obtained in Refs.\onlinecite{Magnusson,Savenko}
(see also supplementary material).

\begin{eqnarray}
\frac{\partial\rho}{\partial t}=\frac{i}{\hbar}\left[\rho;H_0+H_{pol-pol}+H_T\right]+\\
\nonumber+\frac{\delta_{\Delta E}}{\hbar}\left\{2\left(H_R^+\rho
H_R^-+H_R^-\rho
H_R^+\right)\right.-\\
\nonumber-\left.\left(H_R^+H_R^-+H_R^-H_R^+\right)\rho-\rho\left(H_R^+H_R^-+H_R^-H_R^+\right)\right\}+\\
\nonumber
+\frac{1}{2\tau_L}\widehat{L}_{a_L}+\frac{1}{2\tau_D}\widehat{L}_{a_D}+\frac{1}{2\tau_R}\widehat{L}_{a_R}+\frac{1}{2\tau}\widehat{L}_{c}+\frac{P}{2}\widehat{L}_{a_U^+}+\frac{I}{2}\widehat{L}_{c^+}
\label{MasterEquation}
\end{eqnarray}
where $\widehat{L}_A$ is Linblad operator defined by the formula
$\widehat{L}_A=2A\rho A^--A^+A\rho-\rho A^+A$ and $\tau_L$,
$\tau_U$, $\tau_R$ and $\tau$ are lifetimes of lower polaritons,
upper polaritons, polaritons in the reservoir and THz photons, and
$P$ and $I$ are pumping intensities of upper polariton state and
terahertz mode. The delta function $\delta_{\Delta E}$ denotes the
conservation of energy in the process of phonon scattering. The
first line accounts for the coherent processes in the system, the
second and third lines correspond to the phonon- assisted coupling
with incoherent reservoir of the polaritons, the last line
accounts for the pump and the decay.

The equations for the  populations of polariton states and
terahertz photons can be obtained as

\begin{equation}
\partial_tn_i=Tr\left(\widehat{n_i}\frac{\partial\rho}{\partial t}\right)
\end{equation}

Using the mean field approximation, one gets the closed system of
the dynamic equations for the occupancies $n_L=\langle
a_L^+a_L\rangle, n_U=\langle a_U^+a_U\rangle,
n_{R\textbf{k}}=\langle a_{R\textbf{k}}^+a_{R\textbf{k}}\rangle$ and
$n=\langle c^+c\rangle$ connected by the  correlators
$\alpha_{LU}=\langle a_L^+a_Uc^+\rangle,\alpha_{UL}=\langle
a_La_U^+c\rangle = \alpha_{LU}^{\ast}$ (see supplementary material)
\begin{eqnarray}
\partial_tn_L=-2\frac{V_T}{\hbar} \textrm{Im}\left(\alpha_{UL}\right)-\frac{n_L}{\tau_L}+~~~\\
\nonumber
+W_2\sum_{\textbf{k}}\{\left(n_L+1\right)n_{R\textbf{k}}\left(n_{\textbf{k}}^{ph}+1\right)-n_L\left(n_{R\textbf{k}}+1\right)n_{\textbf{k}}^{ph}\};
\end{eqnarray}
\begin{eqnarray}
\partial_tn_U=2\frac{V_T}{\hbar}\textrm{Im}\left(\alpha_{UL}\right)-\frac{n_U}{\tau_U}+P+~~~\\
\nonumber
+W_1\sum_{\textbf{k}}\{\left(n_U+1\right)n_{R\textbf{k}}n_{\textbf{k}}^{ph}-n_U\left(n_{R\textbf{k}}+1\right)\left(n_{\textbf{k}}^{ph}+1\right)\};
\end{eqnarray}
\begin{eqnarray}
\partial_tn_{R\textbf{q}}=-\frac{n_{R\textbf{q}}}{\tau_R}+~~~\\
\nonumber
W_1\sum_{\textbf{k}}\{n_U\left(n_{R\textbf{k}}+1\right)\left(n_{\textbf{k}}^{ph}+1\right)-\left(n_U+1\right)n_{R\textbf{k}}n_{\textbf{k}}^{ph}\}+\\
\nonumber
+W_2\sum_{\textbf{k}}\{n_L\left(n_{R\textbf{k}}+1\right)n_{\textbf{k}}^{ph}-\left(n_L+1\right)n_{R\textbf{k}}\left(n_{\textbf{k}}^{ph}+1\right)\};\\
\nonumber
\end{eqnarray}
\begin{eqnarray}
\partial_tn=-2\frac{V_T}{\hbar}\textrm{Im}\left(a_{UL}\right)-\frac{n}{\tau}+I\\
\nonumber
\end{eqnarray}
\begin{eqnarray}
\partial_t\alpha_{UL}=\frac{i}{\hbar}\left(\tilde{\epsilon_U}-\tilde{\epsilon}_L-\epsilon_T\right)\alpha_{UL}-\frac{a_{UL}}{\tau_{corr}}+~~~\\
\nonumber
+i\frac{V_T}{\hbar}\{\left(n_U+1\right)n_Ln-n_U\left(n_L+1\right)\left(n+1\right)\}+\\
\nonumber
+\{W_1\sum_{\textbf{k}}\left(-n_{R\textbf{k}}-n_{\textbf{k}}^{ph}-1\right)+W_2\sum_{\textbf{k}}\left(n_{R\textbf{k}}-n_{\textbf{k}}^{ph}\right)\}a_{UL}.
\nonumber
\end{eqnarray}
In the above expressions
$\tau_{corr}^{-1}=\tau_L^{-1}+\tau_U^{-1}+\tau_R^{-1}+\tau^{-1}$,
$V_T\approx 1$ $\mu eV$ is a coupling constant between polaritons
and terahertz photons and  $W_{1,2}\approx 2ps^{-1}$ are transition
rates between the reservoir and upper/lower polariton states
determined by polariton-phonon interaction constants, $W_{1,2}\sim|{D_{1,2}}|^2$.
Note, that chatracteristic
time of terzhetz photon emission is about three orders of
magnitude smaller than characteristic time of the scattering with
acoustic phonons. However, THz emission is dramatically inhanced
by bosonic stimulation and becomes dominant mechanism for sufficiently strong pumps.
$n_{\textbf{k}}^{ph}$ gives the occupancies of the phonon mode
determined by Bose distribution function. For simplicity of the
calculations in the present paper we consider the reservoir to
consist from N identical states ($N=3\cdot10^{5}$). Note, that if
coherent interaction is switched off by equating $d\alpha_{UL}/dt=0$ the system of the equations we use transforms
into the system of Boltzmann equations considered in
Ref.\onlinecite{Kavokin1}.

The renormalized energies of the upper and lower polariton states
are determined by their blueshifts arising from polariton-polariton
interactions and read
\begin{eqnarray}
\tilde{\epsilon}_U=\epsilon_U+2\left(U_{UU}n_U+U_{UL}n_L+U_{UR}\sum_\textbf{k}n_{R\textbf{k}}\right),\\
\tilde{\epsilon}_L=\epsilon_L+2\left(U_{LL}n_L+U_{UL}n_U+U_{LR}\sum_\textbf{k}n_{R\textbf{k}}\right).
\end{eqnarray}
Due to the difference of the Hopfield coefficients for the upper and
lower polariton states, the difference $\tilde{\epsilon}_U-\tilde{\epsilon}_L$ depends on the polariton
concentrations and thus is determined by the intensity of the pump
$P$. This dependence can have important consequencies, allowing for
the onset of the bistability in the system (see below).

\begin{figure}[!htb]
\includegraphics[width=1.0\linewidth]{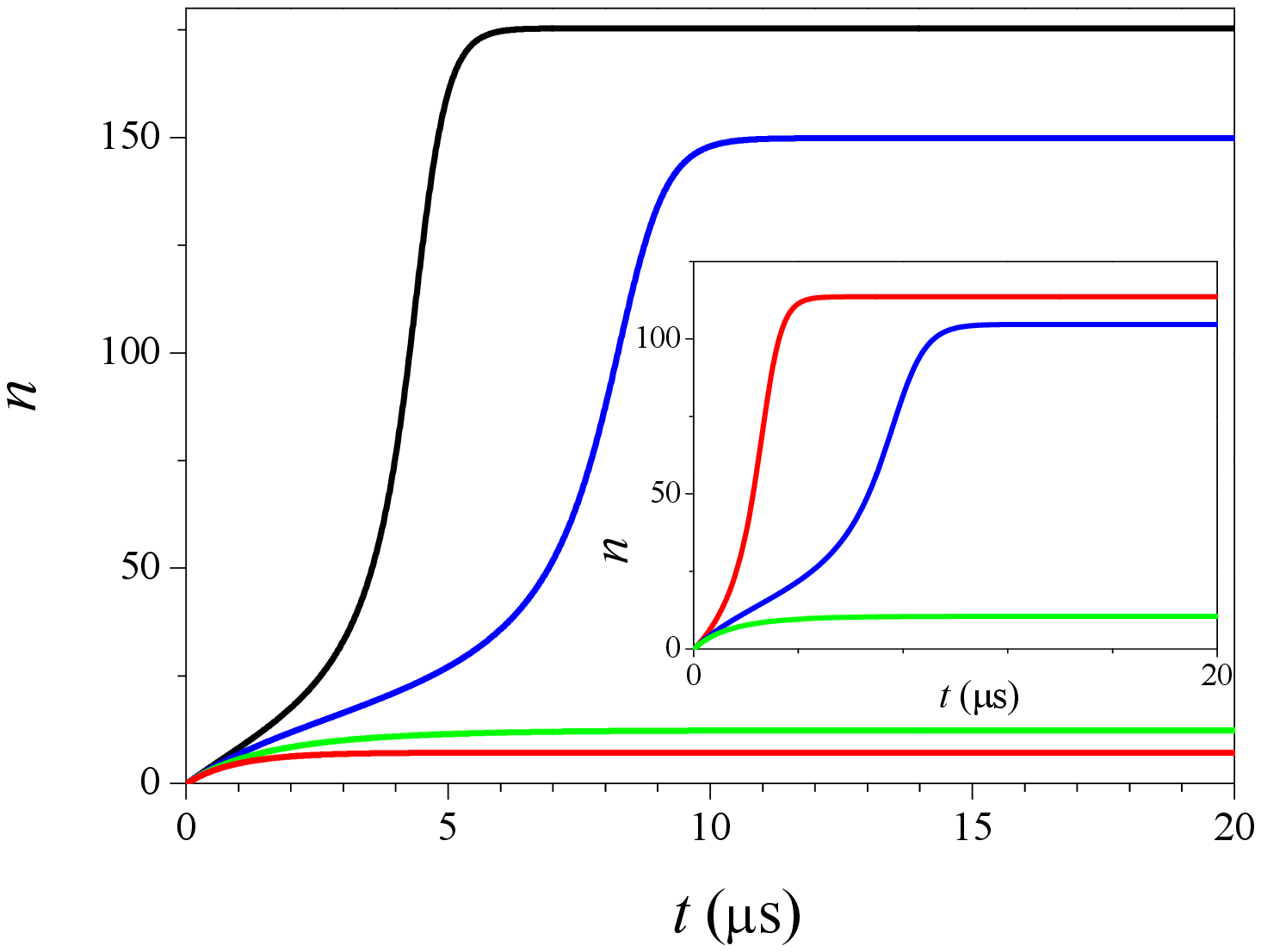}
\caption{Time evolution of terahertz photons number at zero
temperature for different pumps: $4500$ ps$^{-1}$ (red), $5000$
ps$^{-1}$ (green), $5500$ ps$^{-1}$ (blue) and $6000$ ps$^{-1}$
(black). Inset shows evolution of THz photons number for the
constant pump $P=6$ $10^3$ ps$^{-1}$ for different temperatures: 1 K
(red), 10 K (green) and 20 K (blue).} \label{fig:2}
\end{figure}

\emph{Results and discussions}. We consider a planar $GaAs$ microcavity
in strong coupling regime with Rabi splitting $\Omega_R$
between upper and lower polariton modes equal to 16 meV (which
corresponds to 4 THz) and embedded into THz cavity with eigen
frequency slightly different from $\Omega_R$ and having a quality factor
$Q=100$\cite{Chassagneux2,Gallant}. Let us assume that initially the
system is characterized by zero population of polaritons and THz
photons. When the constant non-resonant pump of the upper polariton
state is switched on, the number of THz photons $n$ starts to
increase until it reaches some equilibrium level defined by the
radiative decay of polaritons and escape of THz radiation from the cavity, as it is shown at
Fig.\ref{fig:2}

\begin{figure}[!htb]
\includegraphics[width=1.0\linewidth]{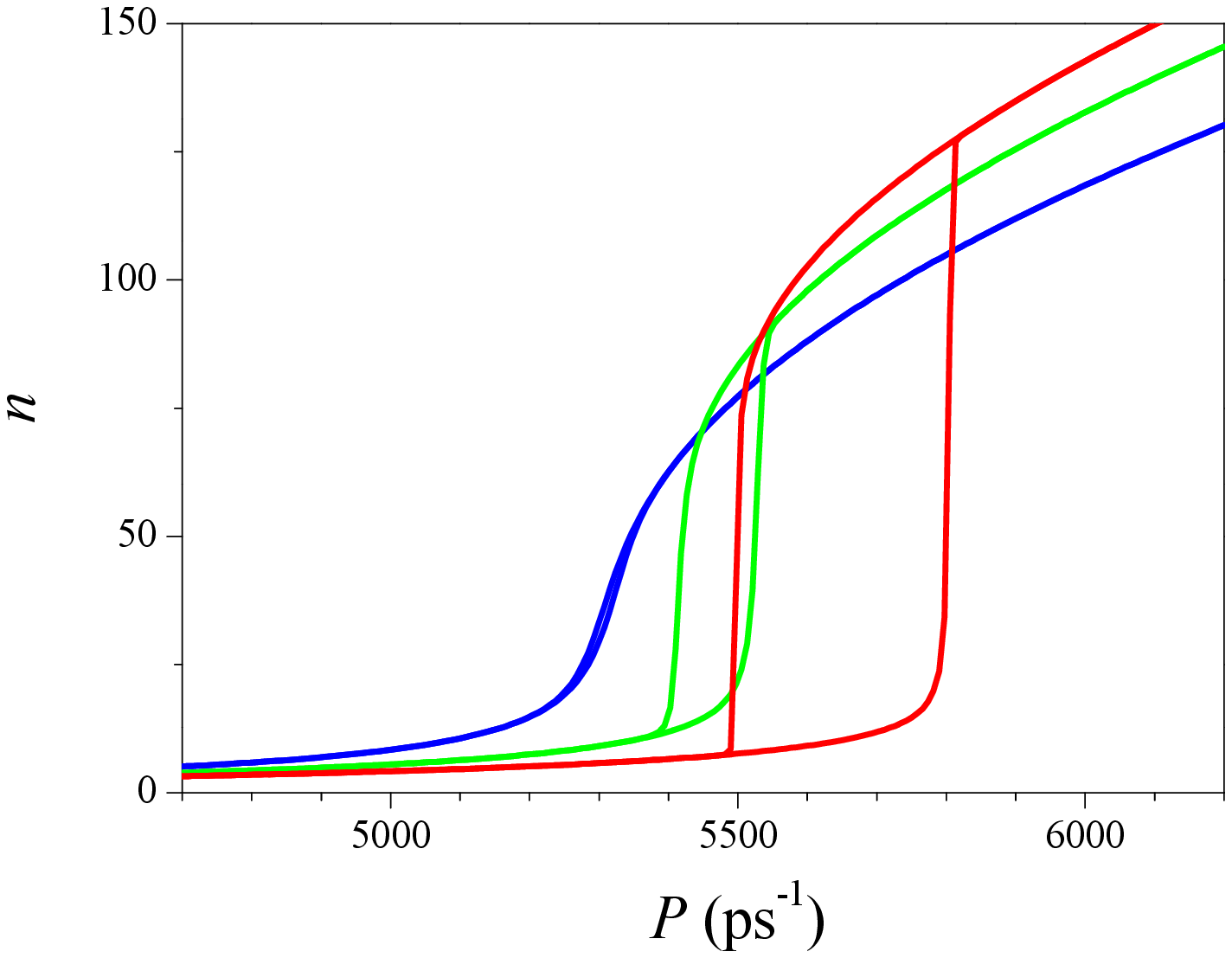} \\
\caption{Dependence of occupancy of the THz mode on pump in
equilibrium state for different temperatures: 1 K (red), 10 K
(green) and 20 K (blue).} \label{fig:3}
\end{figure}

Equilibrium value of the THz population $n$ as a function of pumping $P$ demonstrates threshold-like
behavior. For high enough temperatures, below the threshold the dependence of $n$ on $P$ is
very weak. When pumping reaches the certain threshold value,
polariton condensate is formed in the lower polariton state,
radiative THz transition is amplified by bosonic stimulation, and the
occupancy of THz mode increases superlinearly  together with the
occupancy of lower polariton state $n_L$ (Fig.\ref{fig:2}, blue
curve).  This behavior is qualitatively the same as in the approach
operating with semiclassical Boltzmann equations. However, the decrease
of temperature leads to the onset of the bistability and
hysteresis in the dependence $n(P)$. The bistable jump occurs when
the intensity of the pump tunes
$\tilde{\epsilon}_U-\tilde{\epsilon}_L$ into the resonance with the cavity
mode $\epsilon_T$. The parameters of the hysteresis loop strongly
depend on the temperature (Fig.\ref{fig:3}). It is very
pronounced and broad for low temperatures, narrows with the increase of the temperature
and disappears completely at $T\approx20K$.

\begin{figure}[!htb]
\includegraphics[width=1.0\linewidth]{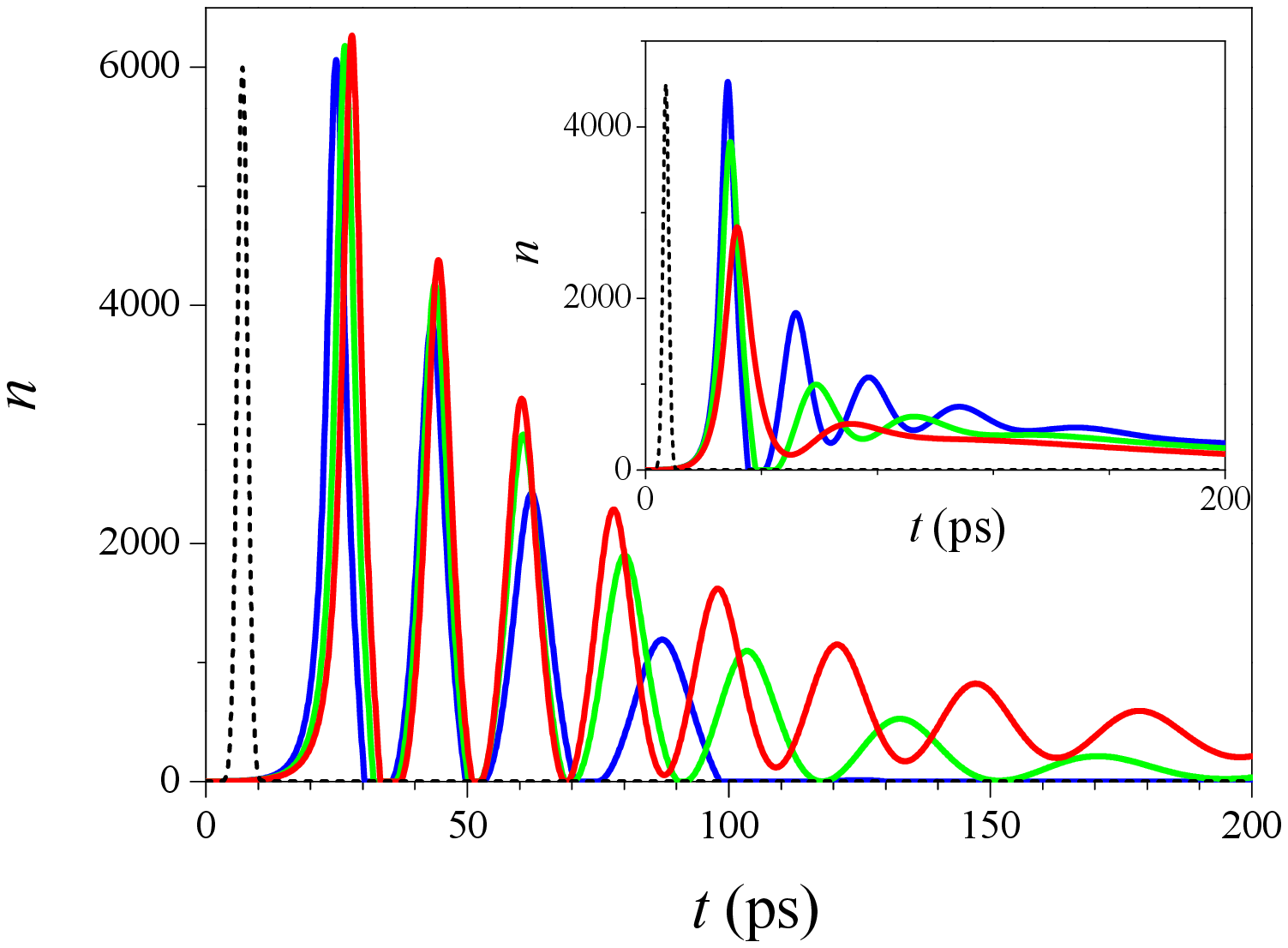}after the arrival of the short excitation pulse (dashed black line).
\caption{The temporal dependence of the terahertz mode occupancy.
The background pump is switched off, lifetimes $\tau_L=\tau_U=50$ ps. The temperatures are: 1 K (red line),
10 K (green) and 20 K (blue).   Inset: T=1K, different lifetimes of the  polariton states: 15 ps
(red), 20 ps(green) and 25 ps(blue).} \label{fig:4}
\end{figure}

Coherent nature of the interaction beween excitons and THz photons
makes possible the periodic exchange of the energy between
polaritonic and photonic modes and oscillatory dependence of the THz
signal in time. Fig.\ref{fig:3} shows temporal evolution of the
occupancy of THz mode after excitation of the upper polariton state
by a short pulse having a duration of about 2 ps. It is seen that the
occupancy of THz mode reveals a sequence of the short pulses having
duration of dozens of ps with amplitude decaying in time due to
escape of THz photons from a cavity and radiative decay of
polaritons. The period of the oscillations is sensitive to the
number of the injected polaritons $N_0$ and decreases with
increasing of $N_0$. If the lifetime of polaritons is less than the
period of the oscillations, single pulse behaviour can be observed
as it is shown in the inset of Fig.\ref{fig:4}. Appropriate choice
of the parameters can ultimately lead to a generation of THz
wavelets composed of one or several THz cycles, which makes
polariton-THz system suitable for application in a sort pulse THZ
spectroscopy.

\begin{figure}[!htb]
\includegraphics[width=1.0\linewidth]{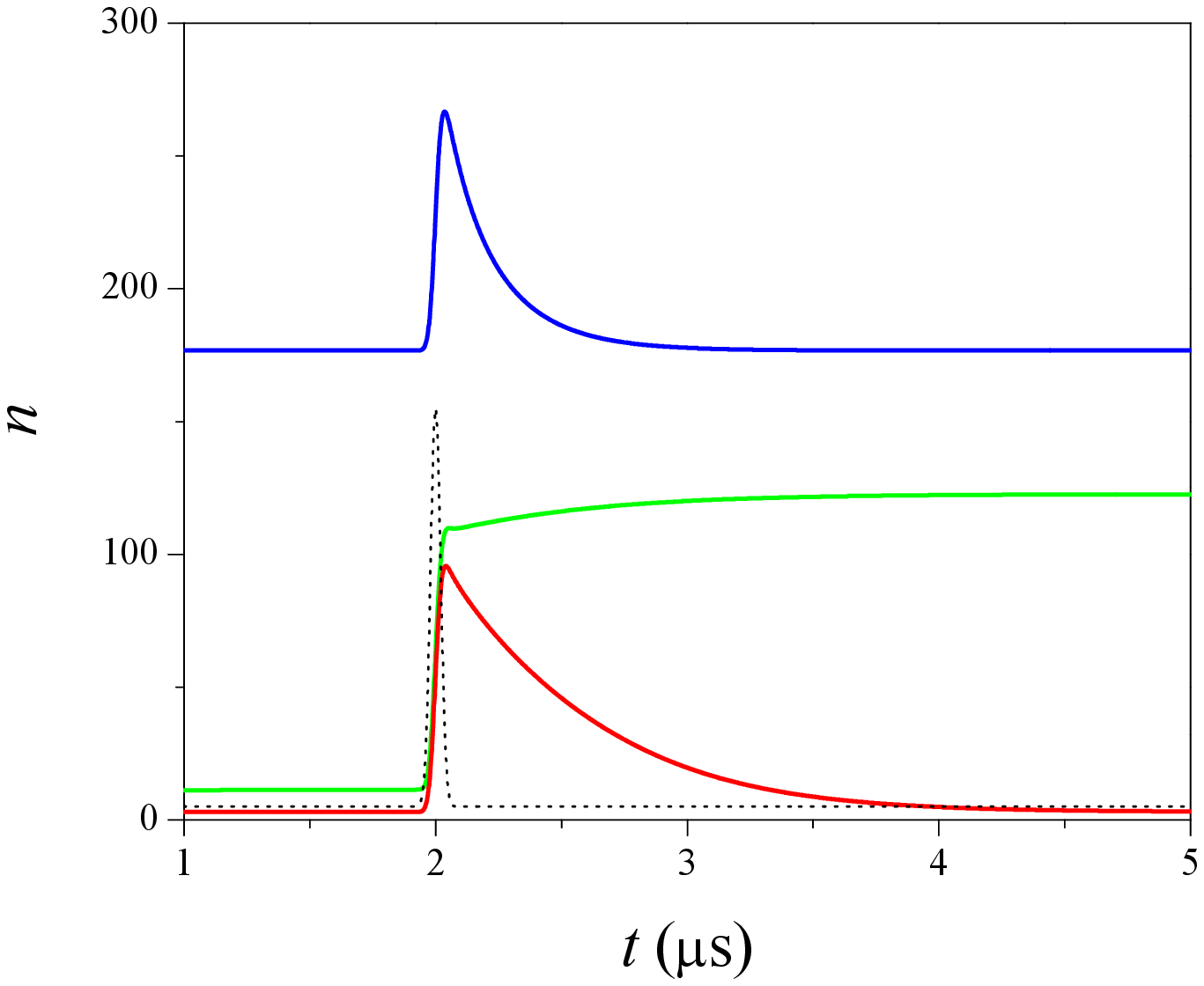}
\caption{System response on a short single impuls (FWHM$=40$ ps,
black/dotted curve) for different values of the background pump:
4600 ps$^{-1}$ (red), 5750 ps$^{-1}$ (green), 6500 ps$^{-1}$ (blue).
Switching occurs only when the background pump corresponds to the
bistable region (compare with Fig. 2).} \label{fig:5}
\end{figure}

If the system of coupled THz photons and cavity polaritons is in the
state corresponding to the lower branch of the S-shaped curve in the
bistability region, illumination of the system by a short THz
impulse $I(t)$ can induce its switching to the upper branch, as it
is demonstrated at Fig.\ref{fig:5}. One sees, that the response of
the system is qualitatively different for different values of the pump $P$.
If $P$ lies outside the bistablity region, the
application of THz pulse leads to a short increase of the $n$ but
subsequently the system relaxes to its
original state (red and blue curves). However, when the system is in
the bistability regime the switching occurs. Note that this effect is of
quantum nature and cannot be describes using the approach based on
semiclassical Boltzmann equations developed in
Ref.\onlinecite{Kavokin1}.


\emph{Conclusion}. We considered the system of coupled cavity
polaritons and THz photons using the approach based on generalized
Lindblad equation for the density matrix. We showed that such
system demonstrates a variety of intriguing nonlinear effects,
including bistability, THz swithching and generation of short THz
wavelets. The work was supported by Rannis "Center of excellence in
polaritonics", FP7 IRSES "POLAPHEN" and "POLALAS" projects, Russian
Fund of Basic Research and COST "POLATOM" program.

\begin{widetext}

\section{Supplementary materials}
In the present supplementary appendix we present a derivation of the quantum kinetic equations for the system of cavity polaritons coupled with a terahertz (THz) cavity mode based on the Lindblad approach for the density matrix dynamics. The method we develop is general and can in principle be applied to any system of interacting bosons in contact with a phonon reservoir, for example, a polaritonic channel [18,19] or a condensate of indirect excitons.

\subsection{The Lindblad approach}
The system of polaritons, phonons and THz cavity photons is described by its density matrix $\rho$, for which we apply Born approximation factorizing it into the phonon part which is supposed to be time-independent and corresponds to the thermal distribution of acoustic phonons $\rho_{ph}=\texttt{exp}\left\{-\beta\widehat{H}_{ph}\right\}$, and the part describing polaritons and THz cavity photons $\rho_{cav}$ whose time dependence should be determined.  $\rho=\rho_{ph}\otimes \rho_{cav}$.  Our aim is to find dynamic equations for the time evolution of the occupancies of the upper and lower polariton states and the THz cavity mode:
\begin{eqnarray}
n_{i}(t)=Tr\left\{\widehat{a}_i^\dagger\widehat{a}_i\rho(t)\right\}
\equiv\langle\widehat{a}_i^\dagger\widehat{a}_i\rangle
\end{eqnarray}
where $\widehat{a}_i^\dagger$ and $\widehat{a}_j$ are operators of the upper and lower polaritons ($i=U$ and $i=L$ respectively) and THz cavity photons ($i=T$). In our consideration we neglect spin of the cavity polaritons since out goal is to find the effects of bistability and switching in the THz emitter and spin degree of freedom is not expected to introduce any qualitatively new physics from this point of view. It should be noted, however, that introduction of spin into the model is straightforward.

The total Hamiltonian of the system can be represented as a sum of two parts
\begin{equation}
\widehat{H}=\widehat{H}^{(1)}+\widehat{H}^{(2)}
\end{equation}
where the first term $\widehat{H}^{(1)}$ describes the "coherent" part of the evolution, corresponding to free polaritons, cavity photons and polariton-polariton interactions, and the second term $\widehat{H}^{(2)}$ corresponds to the dissipative interaction with acoustic phonons. The two terms affect the polariton density matrix in a qualitatively different way. The effect of the coherent part on the evolution of the density matrix is described by the Liouville-von Neumann equation
\begin{equation}
i\left(\partial_t\rho\right)^{(1)}=\left[\widehat{H}^{(1)};\widehat{\rho}\right]
\label{liouville}
\end{equation}

Polariton-phonon scattering corresponds to the interaction of the quantum polariton system with a classical phonon reservoir. It is of
dissipative nature, and thus straightforward application of the Liouville-von Neumann equation is impossible. Introduction of
dissipation into quantum systems is an old problem, for which there is still no single well established solution. In the domain of quantum
optics, however, there are standard methods based on the Master Equation techniques. In the following we give a brief outline of this approach applied for a dissipative polariton system.

The Hamiltonian of interaction of polaritons with acoustic phonons in Dirac representation can be represented as
\begin{eqnarray}
\widehat{H}^{(2)}(t)=\widehat{H}^-(t)+\widehat{H}^+(t)=D_U\sum_{k}e^{i\left(E_U-E_{Rk}\right)t}\left(\widehat{a}_U\widehat{a}_{Rk}^++\widehat{a}_U^+\widehat{a}_{Rk}\right)\left(\widehat{b}_ke^{-i\omega_k t}+\widehat{b}_k^+e^{i\omega_k t}\right)+\\
\nonumber
+D_L\sum_{k}e^{i\left(E_{Rk}-E_L\right)t}\left(\widehat{a}_L\widehat{a}_{Rk}^++\widehat{a}_L^+\widehat{a}_{Rk}\right)\left(\widehat{b}_ke^{-i\omega_k t}+\widehat{b}_k^+e^{i\omega_k t}\right)
\end{eqnarray}
where $\widehat{a}_i$ are the operators for polaritons, $\widehat{b}_k$ are the operators for phonons, $E_i$ and $\omega_k$ are the dispersion relations of polaritons and acoustic phonons respectively, $D_{i}$ are the polariton-phonon coupling constants. In the last equality we separated the terms $\widehat{H}^+$ where a phonon is created, containing the operators $\widehat{b}^+$, from the terms $\widehat{H}^-$ in which it is destroyed, containing operators $\widehat{b}$.

Now, one can consider a hypothetical situation when polariton-polariton interactions are absent, and all redistributions of the polaritons are due to the scattering with a thermal reservoir of acoustic phonons. One can rewrite the Liouville-von Neumann equation in an integro-differential form and apply the so-called Markovian approximation
\begin{eqnarray}
\left(\partial_t\rho\right)^{(2)}=-\frac{1}{\hbar^2}\int_{-\infty}^t\left[\widehat{H}^{(2)}(t);\left[\widehat{H}^{(2)}(t');\widehat{\rho}(t)\right]\right]=\\
\nonumber
=\delta_{\Delta E}\left[2\left(\widehat{H}^+\widehat{\rho} \widehat{H}^-+\widehat{H}^-\widehat{\rho}\right.\widehat{H}^+\right)\left.-\left(\widehat{H}^+\widehat{H}^-+\widehat{H}^-\widehat{H}^+\right)\widehat{\rho}-\widehat{\rho}\left(\widehat{H}^+\widehat{H}^-+\widehat{H}^-\widehat{H}^+\right)\right]=\widehat{L}_{H^{(2)}}\rho\label{Liouville_int},
\end{eqnarray}
where in the last line symbol $\widehat{L}_{H^{(2)}}$ denotes a Lindblad dissipative operator corresponding to the Hamiltonian $\widehat{H}^{(2)}$. The coefficient $\delta_{\Delta E}$ corresponds to the energy conservation and has dimensionality of inverse energy divided by square of a Plank constant. In calculations we estimate $\delta_{\Delta E}$ as being proportional to the inverse broadening of the polaritons states (as it is usually done in calculation of the transition rates in semiclassical Boltzmann equations using Fermi golden rule). For time evolution of the mean value of any arbitrary operator $ \langle \widehat{A}\rangle=Tr(\rho\widehat{A})$ due to scattering with phonons one thus has:
\begin{equation}\label{eqM}
\left\{\partial_t\langle
\widehat{A}\rangle\right\}^{(2)}=\delta_{\Delta E}\left(\langle[\widehat{H}^-;[\widehat{A};\widehat{H}^+]]\rangle+\langle[\widehat{H}^+;[\widehat{A};\widehat{H}^-]]\rangle\right).
\end{equation}
Putting $\widehat{A}=\widehat{a}_i^+\widehat{a}_i$  in this equation we get the contributions to the dynamic equations for the occupancies coming from polariton-phonon interactions, which are nothing more than the standard semi-classical Boltzmann equations describing the  thermalization of a polariton system.

Combined together, coherent and incoherent contributions result in a following master equation for the density matrix:
\begin{equation}
\frac{\partial\rho}{\partial t}=\frac{i}{\hbar}[\widehat{\rho};\widehat{H}^{(1)}]+\widehat{L}_{H^{(2)}}\rho
\end{equation}

\subsection{Coherent part}
Let us consider the coherent part. Hamiltonian here $\widehat{H}^{(1)}=\widehat{H}_0+\widehat{H}_T+\widehat{H}_{pol-pol}$ (let us omit symbols " $\widehat{}$ " over operators hereafter.), where
\begin{equation}
H_0=\epsilon_La_L^+a_L+\epsilon_Ua_U^+a_U+\epsilon_Tc^+c
\end{equation}

is the free-polariton Hamiltonian,

\begin{equation}
H_T=V_T\left(a_U^+a_Lc+a_Ua_L^+c^+\right)
\end{equation}

is the polaritons-to-THz photons interaction term and

\begin{eqnarray}
H_{pol-pol}=U_{LL}a_L^+a_L^+a_La_L+U_{UU}a_U^+a_U^+a_Ua_U+2U_{UL}a_U^+a_Ua_L^+a_L+2U_{UR}\sum_k{a_U^+a_Ua_{Rk}^+a_{Rk}}+2U_{LR}\sum_k{a_L^+a_La_{Rk}^+a_{Rk}}\approx\\
\nonumber
\approx 2\left[U_{LL}(a_L^+a_L\langle a_L^+a_L\rangle)+U_{UU}\left(a_U^+a_U\langle a_U^+a_U\rangle\right)+U_{UL}\left(a_L^+a_L\langle a_U^+a_U\rangle\right)+U_{UL}\left(a_U^+a_U\langle a_L^+a_L\rangle\right)+\right.\\
\nonumber
\left.+U_{UR}\left(a_U^+a_U\langle a_{Rk}^+a_{Rk}\rangle\right)+U_{UR}\left(a_{Rk}^+a_{Rk}\langle a_{U}^+a_{U}\rangle\right)+U_{LR}\left(a_L^+a_L\langle a_{Rk}^+a_{Rk}\rangle\right)+U_{LR}\left(a_{Rk}^+a_{Rk}\langle a_{L}^+a_{L}\rangle\right)\right]\approx\\
\nonumber
\approx 2\left(U_{LL}n_L+U_{UL}n_U+\sum_k{U_{LR}n_{Rk}}\right)a_L^+a_L+2\left(U_{UU}n_U+U_{UL}n_L+\sum_K{U_{UR}n_{Rk}}\right)a_U^+a_U+\\
\nonumber
+2\left(U_{UR}n_U+U_{LR}n_L\right)\sum_k{a_{Rk}^+a_{Rk}}=U_1a_L^+a_L+U_2a_U^+a_U+U_3\sum_k{a_{Rk}^+a_{Rk}};
\end{eqnarray}

is the polariton-polariton scattering Hamiltonian. Coming from the first to the second lines of this expression we used mean-field approximation.
Here we also introduced three new coefficients:

\begin{eqnarray}
U_1=2\left(U_{LL}n_L+U_{UL}n_U+\sum_k{U_{LR}n_{Rk}}\right),\\
U_2=2\left(U_{UU}n_U+U_{UL}n_L+\sum_K{U_{UR}n_{Rk}}\right),\\
U_3=2\left(U_{UR}n_U+U_{LR}n_L\right).
\end{eqnarray}


\textbf{1) Lower and upper polaritons occupancies} $n_L,n_U$,

Consider the dynamic equation for $n_L$ as an example. One has

\begin{eqnarray}
\nonumber
\hbar\left(\partial_tn_L\right)^{(1)}=i\textrm{Tr}\{a_L^+a_L[\rho;H^{(1)}]\};\\
\nonumber\\
\hbar\left(\partial_tn_L\right)_0=i\textrm{Tr}\{a_L^+a_L[\rho;H_0]\}=i\textrm{Tr}\{\rho[\epsilon_La_L^+a_L+\epsilon_Ua_U^+a_U+\epsilon_Tc^+c;a_L^+a_L]\}=0;\nonumber\\
\hbar\left(\partial_tn_L\right)_T=iV_T\textrm{Tr}\{\rho[a_U^+a_Lc+a_Ua_L^+c^+;a_L^+a_L]\}=iV_T\left(a_{UL}-a_{UL}^+\right)=-2V_T\textrm{Im}\left(a_{UL}\right);
\nonumber\\
\hbar\left(\partial_tn_L\right)_{pol-pol}=i\textrm{Tr}\{\rho[U_1a_L^+a_L + U_2a_U^+a_U+U_3\sum_k{a_{Rk}^+a_{Rk}};a_L^+a_L]\}=0.
\nonumber
\end{eqnarray}
Finally,
\begin{equation}
\hbar\left(\partial_tn_L\right)^{(1)}=-2V_T\textrm{Im}\left(a_{UL}\right).
\end{equation}

The equation for the upper polariton occupancy is obtained in a similar way.

\textbf{2) Reservoir} $n_{Rk}$,
\begin{equation}
\hbar\left(\partial_tn_{Rk}\right)^{(1)}=i\textrm{Tr}\{a_{Rk}^+a_{Rk}[\rho;H^{(1)}]\}=0.\\
\end{equation}

\textbf{3) Terahertz cavity occupancy} $n_T$,
\begin{eqnarray}
\nonumber
\hbar\left(\partial_tn_T\right)^{(1)}=i\textrm{Tr}\{c^+c[\rho;H^{(1)}]\};\\
\nonumber\\
\hbar\left(\partial_tn_T\right)_0=i\textrm{Tr}\{c^+c[\rho;H_0]\}=0;
\nonumber\\
\hbar\left(\partial_tn_T\right)_T=iV_T\textrm{Tr}\{\rho[a_U^+a_Uc+a_Ua_L^+;c^+c]\}=iV_T\left(a_{UL}-a_{UL}^+\right)=-2V_T\textrm{Im}\left(a_{UL}\right);
\nonumber\\
\hbar\left(\partial_tn_T\right)_{pol-pol}=i\textrm{Tr}\{\rho[U_1a_L^+a_L+U_2a_U^+a_U+U_3\sum_k{a_{Rk}^+a_{Rk}};c^+c]\}=0.
\nonumber
\end{eqnarray}
Finally,
\begin{equation}
\hbar\left(\partial_tn_T\right)^{(1)}=-2V_T\textrm{Im}\left(a_{UL}\right).\\
\end{equation}

\textbf{4) Correlators} $a_{UL}$,
\begin{eqnarray}
\nonumber
\hbar\left(\partial_ta_{UL}\right)^{(1)}=i\textrm{Tr}\{a_U^+a_Lc[\rho;H^{(1)}]\};\\
\nonumber\\
\hbar\left(\partial_ta_{UL}\right)_0=i\textrm{Tr}\{a_U^+a_Lc[\rho;H_0]\}=i\textrm{Tr}\{\rho[\epsilon_La_L^+a_L+\epsilon_Ua_U^+a_U+\epsilon_Tc^+c;a_U^+a_Lc]\}=i\left(-\epsilon_L+\epsilon_U-\epsilon_T\right)a_{UL};
\nonumber\\
\hbar\left(\partial_ta_{UL}\right)_T=iV_T\textrm{Tr}\{\rho[a_U^+a_Lc+a_Ua_L^+c^+;a_U^+a_Lc]\}=iV_T\left(\langle a_Ua_L^+a_U^+a_Lc^+c\rangle-\langle a_U^+a_La_Ua_L^+cc^+\rangle\right)=
\nonumber\\
=-iV_T\{\left(n_U+1\right)n_Ln_T-n_U\left(n_L+1\right)\left(n_T+1\right)\};
\nonumber\\
\hbar\left(\partial_ta_{UL}\right)_{pol-pol}=i\textrm{Tr}\{\rho[U_1a_L^+a_L+U_2a_U^+a_U+U_3\sum_k{a_{Rk}^+a_{Rk}};a_U^+a_Lc]\}=
\nonumber\\
=iU_1\textrm{Tr}\{\rho[a_L^+a_L;a_U^+a_Lc]\}+iU_2\textrm{Tr}\{[a_U^+a_U;a_U^+a_Lc]\}=iU_1\left(-a_{UL}\right)+iU_2a_{UL}.
\nonumber
\end{eqnarray}

Finally,
\begin{eqnarray}
\hbar\left(\partial_ta_{UL}\right)^{(1)}=i\left(\epsilon_U-\epsilon_L-\epsilon_T\right)a_{UL}+2V_T\{\left(n_U+1\right)n_Ln_T-\\
\nonumber
n_U\left(n_L+1\right)\left(n_T+1\right)\}+i\{2U_{UU}n_U+U_{UL}\left(n_L-n_U\right)-2U_{LL}n_L+\left(U_{UR}-U_{LR}\right)\sum_k{n_{Rk}}\}a_{UL};
\end{eqnarray}


\subsection{Decoherent part}

To get explicit expresions for the dynamics of $n_L$, $n_U$, $n_{Rk}$, $n_T$ and $a_{UL}$
due to decoherent processes of interaction with the reservoir let us consider
Liouville-von Neumann equation for the density matrix after the Born-Markov approximation Eq.~(6) and the simplest case when only three states $L$, $U$ and $R_k$ are present.

In this case, leaving energy- conserving terms only one gets
\begin{eqnarray}
H^+=D_1\sum_ka_Ua_{kR}^+b_k^++D_{2}\sum_ka_L^+a_{kR}b_k^+\\
H^-=D_1\sum_ka_U^+a_{kR}b_k+D_{2}\sum_ka_La_{kR}^+b_k
\end{eqnarray}

The application of Eq.~(6) gives the following results:\\

\textbf{1) Lower and upper branch polariton occupancies} $n_L,n_U$,

For $n_L$ one has:

\begin{eqnarray}
\nonumber
\left(\partial_tn_L\right)^{(2)}=\delta_{\Delta E}\left([H^-;[a_L^+a_L;H^+]]+[H^+;[a_L^+a_L;H^-]]\right)=2\delta_{\Delta E}[H^+;[a_L^+a_L;H^-]];\\
\nonumber
[a_L^+a_L;H^-]=[a_L^+a_L;D_1\sum_ka_U^+a_{Rk}b_k+D_2\sum_ka_La_{Rk}^+b_k]=-D_2\sum_ka_La_{Rk}^+b_k;\\
\nonumber
[H^+;[a_L^+a_L;H^-]]=[D_1\sum_ka_Ua_{Rk}^+b_k^++D_2\sum_ka_L^+a_{Rk}b_k^+;-D_2\sum_ka_La_{Rk}^+b_k]=\\
\nonumber
[D_2\sum_ka_L^+a_{Rk}b_k^+;-D_2\sum_ka_La_{Rk}^+b_k]=D_2^2\sum_k\{\left(a_L^+a_L+1\right)a_{Rk}^+a_{Rk}\left(b_k^+b_k+1\right)-a_L^+a_L\left(a_{Rk}^+a_{Rk}+1\right)b_k^+b_k\};\\
\nonumber
\end{eqnarray}

Finally,
\begin{eqnarray}
\hbar^2(\partial_tn_L)^{(2)}=W_2\sum_k\{\left(n_L+1\right)n_{Rk}\left(n_k^{ph}+1\right)-n_L\left(n_{Rk}+1\right)n_k^{ph}\},
\end{eqnarray}

where we introduced $W_2=2\delta_{\Delta E}D_2^2$.\\

The equation for $n_U$ is easily obtained in an analogical way.

\textbf{2) Reservoir} $n_{Rk}=\langle a_{Rk}^+a_{Rk} \rangle$,
\begin{eqnarray}
\nonumber
\left(\partial_tn_{Rk}\right)^{(2)}=2\delta_{\Delta E}[H^+;[a_{Rk}^+a_{Rk};H^-]];\\
\nonumber
[a_{Rk}^+a_{Rk};H^-]=[a_{Rk}^+a_{Rk};D_1\sum_ka_U^+a_{Rk}b_k+D_2\sum_ka_La_{Rk}^+b_k]=-D_1\sum_ka_U^+a_{Rk}b_k+D_2\sum_ka_La_{Rk}^+b_k;\\
\nonumber
[H^+[a_{Rk}^+a_{Rk};H^-]]=[D_1\sum_ka_Ua_{Rk}^+b_k^++D_2\sum_ka_L^+a_{Rk}b_k^+;-D_1\sum_ka_U^+a_{Rk}b_k+D_2\sum_ka_La_{Rk}^+b_k]\approx\\
\nonumber
\approx D_1^2\sum_k\{a_U^+a_U\left(a_{Rk}^+a_{Rk}+1\right)\left(b_k^+b_k+1\right)-\left(a_U^+a_U+1\right)a_{Rk}^+a_{Rk}b_k^+b_k\}+\\
\nonumber
+D_2^2\sum_k\{a_L^+a_L\left(a_{Rk}^+a_{Rk}+1\right)b_k^+b_k-\left(a_L^+a_L+1\right)a_{Rk}^+a_{Rk}\left(b_k^+b_k+1\right)\},\\
\nonumber
\end{eqnarray}
where coming between the third and the fourth lines we neglected the off- diaginal elements of the phonon density matrix, supposing $\langle b_k^+b_{k'}\rangle=n^{ph}_k\delta_{k,k'}$

Finally,
\begin{eqnarray}
(\partial_tn_{Rk})^{(2)}\approx W_1\sum_k\{n_U\left(n_{Rk}+1\right)\left(n_k^{ph}+1\right)-\left(n_U+1\right)n_{Rk}n_k^{ph}\}+\\
\nonumber
+W_2\sum_k\{n_L\left(n_{Rk}+1\right)n_k^{ph}-\left(n_L+1\right)n_{Rk}\left(n_k^{ph}+1\right)\}.\\
\nonumber
\end{eqnarray}

\textbf{3) Terahertz cavity occupancy} $n_T=\langle c^+c \rangle$,
\begin{equation}
\left(\partial_tn_T\right)^{(2)}=\delta_{\Delta E}\left([H^-;[c^+c;H^+]]+[H^+;[c^+c;H^-]]\right)=0.\\
\end{equation}

\textbf{4) Decoherent part of the correlator} $a_{UL}=\langle a_U^+a_Lc \rangle$,
\begin{eqnarray}
\nonumber
[a_U^+a_Lc;H^-]=0;\\
\nonumber
[a_U^+a_Lc;H^+]=[a_U^+a_Lc;D_1\sum_ka_Ua_{Rk}^+b_k^++D_2\sum_ka_L^+a_{Rk}b_k^+]=-\sum_kD_1a_{Rk}^+a_Lcb_k^++D_2\sum_ka_U^+a_{Rk}cb_k^+;\\
\nonumber
[H^-;[a_U^+a_Lc;H^+]]=[D_1\sum_ka_U^+a_{Rk}b_k+D_2\sum_ka_La_{Rk}^+b_k;-D_1\sum_ka_{Rk}^+a_Lcb_k^++D_2\sum_ka_U^+a_{Rk}cb_k^+]\approx\\
\nonumber
\approx D_1^2\sum_k\{a_{Rk}^+a_{Rk}\left(a_U^+a_Lc\right)b_k^+b_k-\left(a_{Rk}^+a_{Rk}+1\right)a_U^+a_Lc\left(b_k^+b_k+1\right)\}+\\
\nonumber
+D_2^2\sum_k\{a_{Rk}^+a_{Rk}a_U^+a_Lc\left(b_k^+b_k+1\right)-\left(a_{Rk}^+a_{Rk}+1\right)a_U^+a_Lcb_k^+b_k\}=\\
\nonumber
=D_1^2\sum_k\{-a_{Rk}^+a_{Rk}a_U^+a_Lc-a_U^+a_Lc\left(b_k^+b_k+1\right)\}+D_2^2\sum_k\{a_{Rk}^+a_{Rk}a_U^+a_Lc-a_U^+a_Lcb_k^+b_k\}.\\
\nonumber
\end{eqnarray}

Finally,
\begin{equation}
(\partial_ta_{UL})^{(2)}=\frac{W_1}{2}\sum_k\left(-n_{Rk}-n_k^{ph}-1\right)a_{UL}+\frac{W_2}{2}\sum_k\left(n_{Rk}-n_k^{ph}\right)a_{UL}.\\
\end{equation}

After merging the equations for the coherent processes with the equations for the incoherent phonon-scattering processes and adding finite lifetimes, background and impulse pumps (see main text of the Letter) we solve this self-consistent set of equations and eventually find the evolution of the THz photons occupancy.

\end{widetext}

\end{document}